\documentclass[manuscript]{aastex}

\shorttitle{Dispersal of GBPs at different longitudinal magnetic field strengths}
\shortauthors{Yang et al.}

\begin{document}


\title{Dispersal of G-band bright points at different longitudinal magnetic field strengths}


\author{Yunfei Yang\altaffilmark{1,2,3}, Kaifai Ji\altaffilmark{1}$^{\dagger}$, Song Feng\altaffilmark{1,2,3}, Hui Deng\altaffilmark{1}, Feng Wang\altaffilmark{1,4}, and Jiaben Lin\altaffilmark{2}}
\email{jikaifan@cnlab.net}

\altaffiltext{1}{Faculty of Information Engineering and Automation / Yunnan Key Laboratory of Computer Technology Application, Kunming University of Science and Technology, Kunming 650500, Yunnan, China}
\altaffiltext{2}{Key Laboratory of Solar Activity, National Astronomical Observatories, Chinese Academy of Sciences, Beijing 100012, China}
\altaffiltext{3}{Key Laboratory of Modern Astronomy and Astrophysics, Nanjing University, Ministry of Education, Nanjing 210093, China}
\altaffiltext{4}{Yunnan Observatory, Chinese Academy of Sciences, Yunnan 650011, China}

\renewcommand{\thefootnote}{}
\footnotetext{$^{\dagger}$ Corresponding author}


\begin{abstract}
G-band bright points (GBPs) are thought to be the foot-points of magnetic flux tubes. The aim of this paper is to investigate the relation between the diffusion regimes of GBPs and the associated longitudinal magnetic field strengths. Two high resolution observations of different magnetized environments were acquired with $Hinode$ /Solar Optical Telescope. Each observation was recorded simultaneously with G-band filtergrams and Narrow-band Filter Imager (NFI) Stokes $I$ and $V$ images. GBPs are identified and tracked automatically, and then categorized into several groups by their longitudinal magnetic field strengths, which are extracted from the calibrated NFI magnetograms using a point-by-point method. The Lagrangian approach and the distribution of diffusion indices (DDI) approach are adopted separately to explore the diffusion regime of GBPs for each group. It is found that the values of diffusion index and diffusion coefficient both decrease exponentially with the increasing longitudinal magnetic field strengths whichever approach is used. The empirical formulas deduced from the fitting equations are proposed to describe these relations. Stronger elements tend to diffuse more slowly than weak elements, independently of the magnetic flux of the surrounding medium. This may be because the magnetic energy of stronger elements is not negligible compared with the kinetic energy of the gas and therefore the flows cannot perturb them so easily.
\end{abstract}





\section{INTRODUCTION}

Magnetic fields are ubiquitous in the photosphere and interact with convective plasmas. Photospheric convective plasma flows advect magnetic fields, and eventually concentrate them. This interaction may cause magnetic reconnections and the excitation of magnetohydrodynamic waves, which could contribute to the heating of the solar corona \citep{Alfven47, Parker98, Hughes03, DePontieu07, Tomczyk07, vanBallegooijen11, Stangalini13a, Stangalini13b, Giannattasio14}. Thus, it is important to investigate the interaction between small-scale magnetic elements and plasma flows \citep{Hughes03, Viticchie06}. Given that magnetic fields are passively transported by plasma flows, the motion of magnetic elements is a manifestation of intrinsic organizing processes and can be described in terms of diffusion.

Bright points in the photosphere are thought to be the foot-points of the magnetic flux tubes that the convective motions of granules push violently into intergranular lanes \citep{Stenflo85, Solanki93}. G-band observations are suitable for studying bright points because they appear brighter due to a reduced abundance of the CH-lines molecule at a higher temperature \citep{Steiner01}, and are often referred to as G-band bright points (hereafter GBPs). Therefore, it is possible to measure the dynamics of photospheric magnetic flux tubes, although only a fraction of magnetic elements are thought to be associated with GBPs by observations, magneto-hydrodynamical simulations, or semi-empirical models of flux concentrations \citep{Keller92, Berger01, Sanchez01, Steiner01, Schussler03, Carlsson04, Shelyag04, Beck07, Ishikawa07, deWijn08}.

In the past decades, many authors focused on the photospheric dispersal of magnetic elements or bright points in a field of view (FOV), such as an active region (AR) or a quiet Sun (QS) region. Diffusion processes represent the efficiency of dispersal in the photosphere, which uses a diffusion index, $\gamma$, quantifying the transport process with respect to a normal diffusion (random walk). Historically, normal-diffusion ($\gamma$\,=\,1) is the first known diffusion process. It characterizes a trajectory that consists of successive random steps and is described by a simplest form of diffusion theory \citep{Fick55, Einstein05, Lemons97}. \citet{Leighton64} first held that the dispersal rate of magnetic regions in the photosphere is normal. Later, \citet{Jokipii68} and \citet{Muller94} suggested that photospheric magnetic elements move in a random walk due to supergranular flows.
When $\gamma$\,$\neq$\,1, the process is termed anomalous diffusion. The motion of magnetic elements in ARs or network regions in the QS is sub-diffusive ($\gamma$\,$<$\,1) due to being trapped at stagnation points (i.e. points with nearly to zero horizontal velocity; sinks of flow field; \citealt{Lawrence93, Simon95, Cadavid99}).
More evidence based on high resolution data show that the photospheric BP motion is probably super-diffusive ($\gamma$\,$>$\,1; \citealt{Berger98, Lawrence01, Keys14, Yang15}. Typically, \citet{Abramenko11} indicated that the $\gamma$ value increases from a plage area in AR ($\gamma$\,=\,1.48) to a QS region ($\gamma$\,=\,1.53), and to a coronal hole (CH; $\gamma$\,=\,1.67).

The diffusion coefficient, $K$, expresses the rate of increase in the dispersal area in units of time for magnetic elements or GBPs. The $K$ values range from 60 to 176\,$\rm km^{2}$\,$\rm s^{-1}$ in ARs or network regions in the QS \citep{Wang98, Schrijver90, Berger98, Hagenaar99, Giannattasio14, Keys14}, and from 190 to 400\,$\rm km^{2}$\,$\rm s^{-1}$ in QS regions \citep{Schrijver90, Berger98, Giannattasio14, Jafarzadeh14, Yang15}. In particular, \citet{Abramenko11} compared the $K$ values in an AR and QS region, which were about 12 and 19\,$\rm km^{2}$\,$\rm s^{-1}$, respectively, whereas it was 22\,$\rm km^{2}$\,$\rm s^{-1}$ in a CH region.

However, studies focusing on the diffusion of GBPs at different magnetic field strengths are scarce.
Recent high spatial and high temporal magnetograms acquired with the $Hinode$ /Solar Optical Telescope (SOT: \citealt{Kosugi07, Ichimoto08, Suematsu08}) provide an unprecedented opportunity to map a GBP to its co-spatial magnetic field strength. For instance, \citet{Utz13} extracted the corresponding magnetic field strength of each GBP from the longitudinal magnetogram within the spectro-polarimeter (SP) data. Nevertheless, high temporal SP data have such a narrow FOV that they are unsuitable for tracking the complete trajectories of GBPs. Instead, Stokes $I$ and $V$ images of Narrow-band Filter Imager (NFI) are co-spatial and co-temporal with G-band images, and keep a large FOV in observations. Thus, it is feasible to extract the simultaneous longitudinal magnetic field strength of each GBP during its lifetime.
The aim of this paper is to study the relation between the dispersal of GBPs and the associated longitudinal magnetic field strengths. It will shed light on how magnetic elements of different longitudinal fields diffuse on the solar surface, and then assist the study of interaction between convections and magnetic fields.

The layout of the paper is as follows. The observations and data sets are described in Section 2. The data reduction and analysis are detailed in Section 3. In Section 4, the relation of diffusions of GBPs on different longitudinal magnetic field strengths is presented. Finally, the discussion and conclusion are given in Section 5 and 6, respectively.

\section{DATA SETS}
We used two data sets of different magnetized environments acquired with Hinode /SOT. Each data set comprises G-band filtergrams (BFI) and Stokes $I$ and $V$ images (NFI). The G-band data with a wavelength of 4305 {\AA} are suitable for photospheric bright point sensitive investigations. The circular polarization $I$ and $V$ images can measure longitudinal magnetic fields.

In data set \uppercase\expandafter{\romannumeral1}, there is an AR (NOAA 10960) between 2007 May 30 and June 14. The region produced numerous C- and M-class flares during that period. We adopted a timeseries of G-band images taken between 22:31:21 and 23:37:26\,$\rm UT$ on June 9. The images were obtained at a 30\,$\rm s$ cadence with an exposure time of 0.15\,$\rm s$. The spatial sampling is 0.109$''$ over a 111.6$''\times$111.6$''$ FOV. The center of FOV is $x$\,=\,429.5$''$ and $y$\,=\,-158.5$''$ corresponding to a heliocentric angle of 28.2$^{\circ}$ equaling a cosine value of 0.88. We also used a co-spatial and co-temporal timeseries of Stokes $I$ and $V$ images taken 200 m{\AA} blueward of the Na \textsc{i} D 5896{\AA} spectral line with a spatial sampling of 0.16$''$ over a 327.7$''\times$163.8$''$ FOV.

Data set \uppercase\expandafter{\romannumeral2} was recorded on 2007 March 31. It covers a quiet Sun region near the solar disc center. The FOV of the G-band data is 111.6$''\times$55.8$''$ with a resolution of 0.109$''$, and its center pointed to solar coordinates of $x$\,=\,207.7$''$ and $y$\,=\,-130.3$''$. This position corresponds to a heliocentric angle of 12.4$^{\circ}$ equaling a cosine value of 0.98. The data set starts from 11:36\,$\rm UT$ until 12:40\,$\rm UT$, with a temporal sampling of 35\,$\rm s$. The timeseries of co-spatial and co-temporal Stokes $I$ and $V$ images taken 120 m{\AA} blueward of Fe \textsc{i} 6302.5{\AA} spectral line has an FOV of 163.8$''\times$81.9$''$ with a resolution of 0.16$''$.

\section{DATA REDUCTION AND ANALYSIS}
The filtergrams were calibrated and reduced to level-1 using a standard data reduction algorithm fg\_prep.pro. The projection effects of both data sets were corrected according to the heliocentric longitude and latitude of each pixel.

\subsection{Alignment}
The G-band images and the Stokes $I$ and $V$ images was aligned carefully. The temporally closest images were chosen first, and the different spatial sampling were overcome by bicubic interpolating the Stokes $I$ and $V$ images to the corresponding spatial sampling of the G-band images. After that, a sub\_pixel level image registration procedure \citep{Feng12, Yang15} was employed for spatial alignment: All G-band images in the timeseries were aligned to its first image, and then the Stokes $I$ and $V$ images were aligned to the G-band images according to the displacement between the simultaneous Stokes $I$ image and the G-band image. Finally, all of the images were cut for the same FOV.

\subsection{Calibration of NFI Magnetogram}
The NFI Stokes $I$ and $V$ images of two data sets were measured with Na \textsc{i} D 5896{\AA} and Fe \textsc{i} 6302.5{\AA} spectral lines, respectively. Calibration of the NFI magnetograms is needed because they do not allow a full Stokes inversion to be performed. We calibrated the NFI longitudinal magnetic field strengths with a reference to the longitudinal magnetograms in the level-2 SP data, which is inverted from the MERLIN code \citep{Lites07}. As indicated in the previous studies for magnetic fields outside sunspots, a linear relation, $B_{\parallel} = \beta V/I$, can be calibrated between the circular polarization, $V/I$, and the longitudinal field, $B_{\parallel}$ in weak-field approximation \citep{Jefferies89, Chae07, Ichimoto08, Zhou10}.

The NFI $V/I$ images were desampled to consist with the spatial sampling of the SP longitudinal magnetogram. Then the narrow stripes of the timeseries NFI $V/I$ images were cut and aligned with a reference to the co-spatial and co-temporal stripes of the SP longitudinal magnetograms by image registration techniques. The SP observation of data set \uppercase\expandafter{\romannumeral2} was simultaneous with the NFI images of Fe spectral line, so the calibrated coefficient of data set \uppercase\expandafter{\romannumeral2} was measured using the simultaneously SP data directly. Unfortunately, co-spatial and co-temporal SP data were not taken with the NFI images of data set \uppercase\expandafter{\romannumeral1}, so another NFI observation (the same as Na spectral line of data set \uppercase\expandafter{\romannumeral1}, which had simultaneously SP data), was employed to measure the $\beta$ value of Na line. This observation was recorded from an AR between 20:20:05 and 20:59:29\,$\rm UT$ on 2007 July 1, which was temporally close to data set \uppercase\expandafter{\romannumeral1}. Both SP data were obtained by scanning a narrow slit over an area of 297.1$''\times$163.8$''$ with a resolution of 0.30$''\times$0.32$''$ per pixel.

\begin{figure}
\epsscale{1.1}
\plottwo{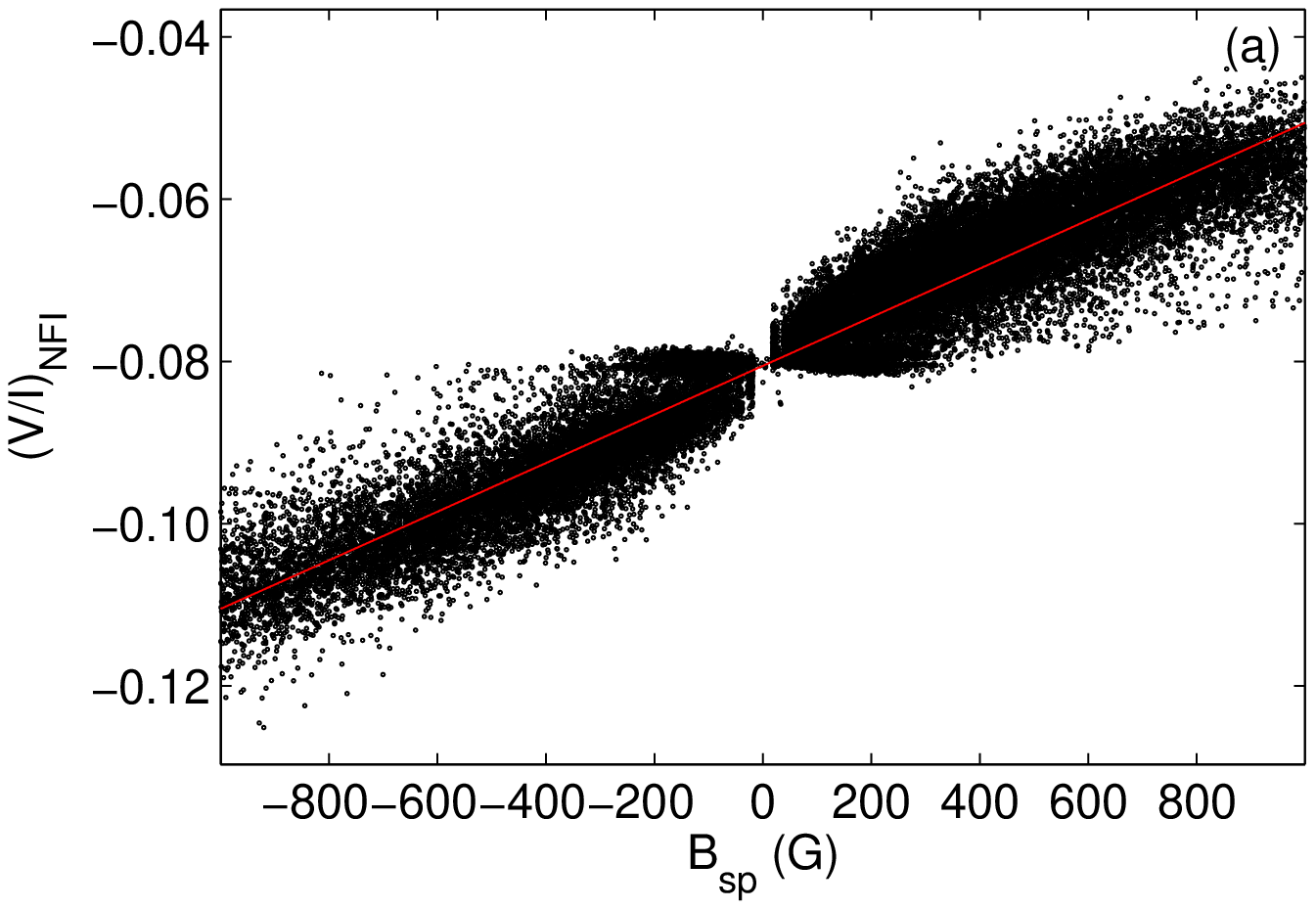}{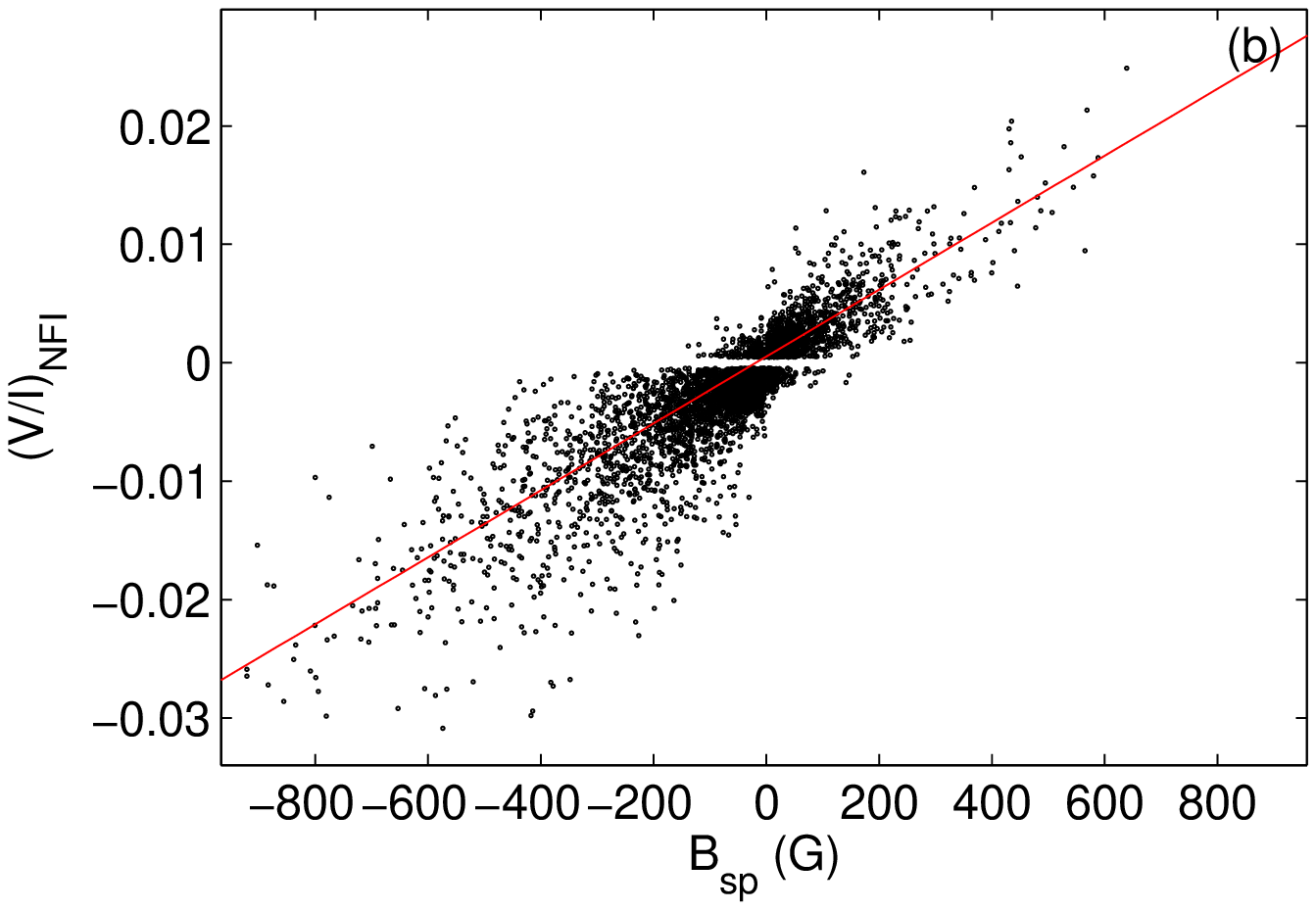}
\caption{Plots of NFI $V/I$ versus $B_{sp}$ of different spectral lines. Panel (a): plot of NFI $V/I$ versus $B_{sp}$ of 200 m{\AA} blueward of the Na \textsc{i} D 5896{\AA} spectral line. The data were recorded between 20:20:05 and 20:59:29\,$\rm UT$ on 2007 July 1. Panel (b): plot of NFI $V/I$ versus $B_{sp}$ of data set \uppercase\expandafter{\romannumeral2}, with 120 m{\AA} blueward of the Fe \textsc{I} 6302.5{\AA} spectral line spectral line. See the electronic edition of the Journal for a color version of this figure.\label{fig1}}
\end{figure}

Figure~\ref{fig1} shows the plots of NFI $V/I$ versus $B_{sp}$ of Na and Fe spectral lines, respectively. The SP longitudinal magnetograms were multiplied by corresponding filling factors. The pixels where the values were less than noise of SP longitudinal magnetograms or NFI $V/I$ images were excluded. Only the pixels whose absolute field strengths are less than 1000\,$\rm G$ were taken into account in data set \uppercase\expandafter{\romannumeral1}. With linear regression analyses of the relations of Na and Fe spectral lines, the $\beta$ values are 13.1 and 35.6\,$\rm kG$ with correlation coefficients of 0.95 and 0.87, respectively. Consequently, longitudinal magnetic fields at any location in the NFI data of data set \uppercase\expandafter{\romannumeral1} and \uppercase\expandafter{\romannumeral2} were determined by multiplying the $V/I$ with the corresponding $\beta$ values. The mean longitudinal magnetic field strengths of data set \uppercase\expandafter{\romannumeral1} (excluding the sunspots region) and \uppercase\expandafter{\romannumeral2} are 132 and 64\,$\rm G$, respectively. In addition, we selected a relatively quiet region (about 1$''\times$1$''$) in the NFI magnetograms to quantify the noise of NFI magnetograms. The standard deviations ($\sigma$) are 10 and 20\,$\rm G$, respectively.

For estimating the variation of the calibrated coefficients in different observations, we also processed some other data sets observed with Na \textsc{i} D 5896{\AA} lines. The result is that their $\beta$ values are stable and limited in the range of 10\%. Therefore, it is feasible to calibrate the NFI magnetograms of data set \uppercase\expandafter{\romannumeral1} using the calibration coefficient of the Na spectral line measured by another observation.

\subsection{Detecting and Tracking GBPs}
\begin{figure}
\epsscale{1.1}
\plottwo{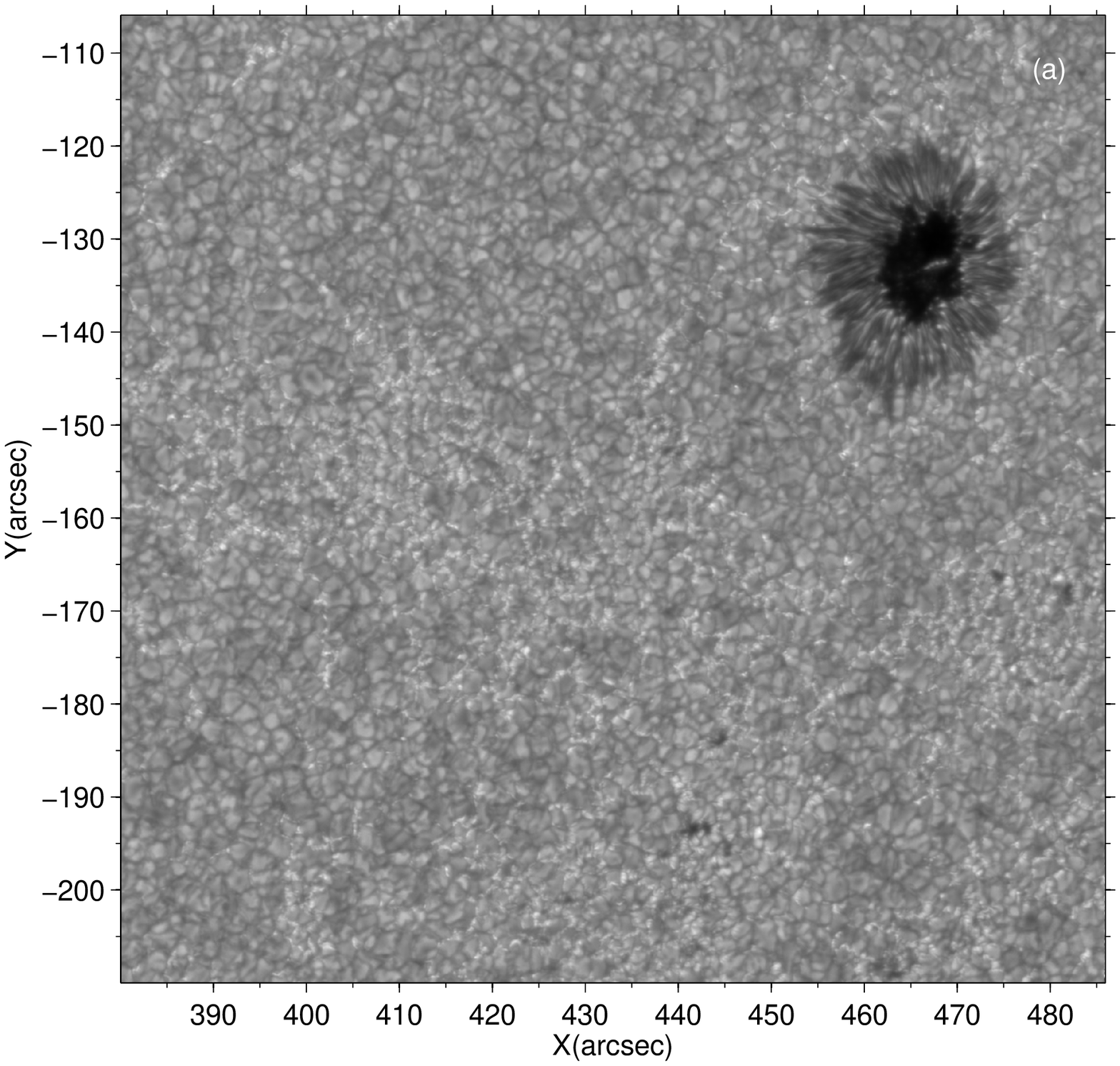}{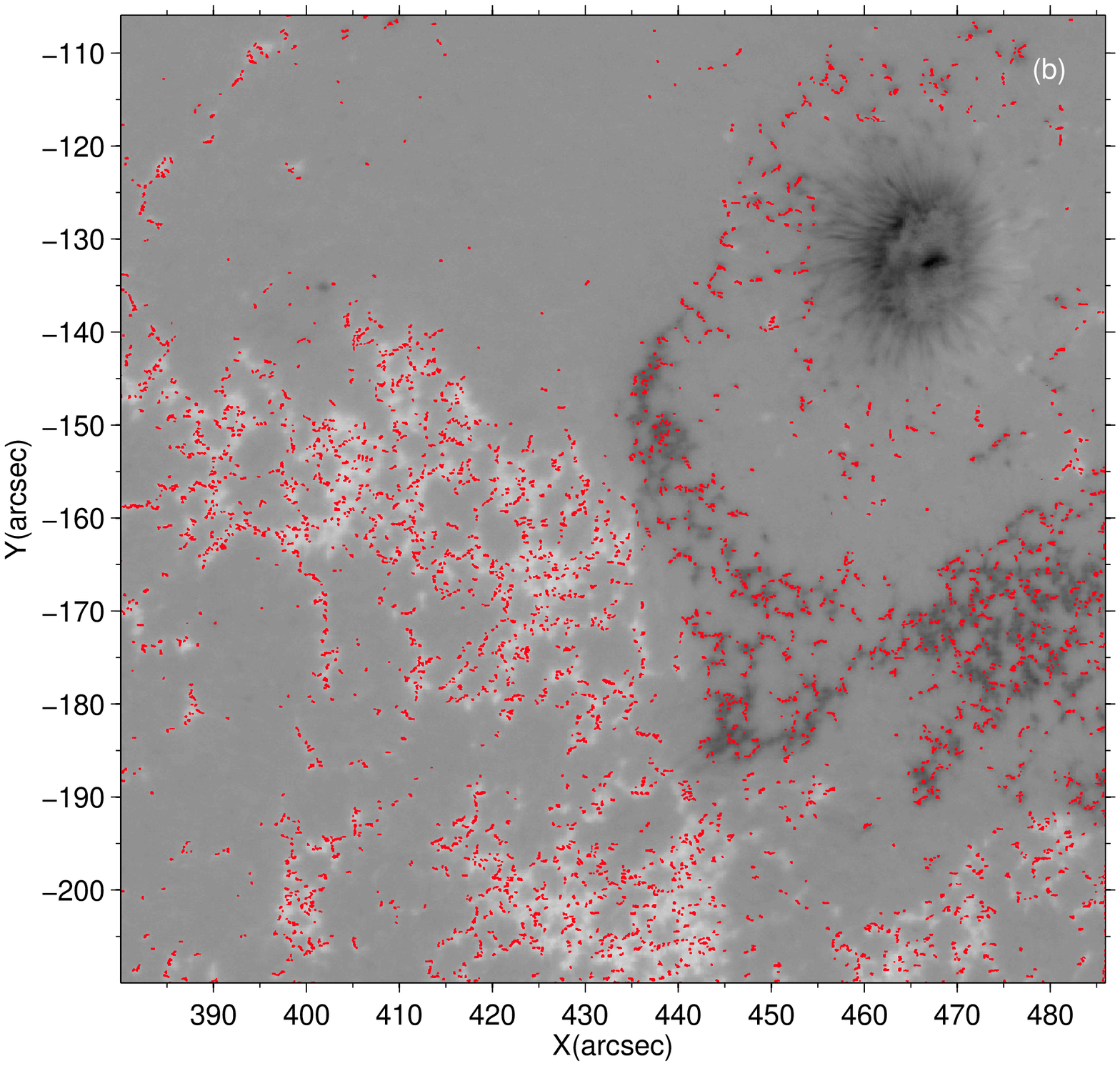}
\caption{Panel (a): a G-band image of data set \uppercase\expandafter{\romannumeral1} (NOAA 10960), which the size is 104.1$''\times$105.9$''$. Panel (b): the co-aligned NFI magnetogram; the GBPs identified from the G-band image are highlighted with red. The $x$ and $y$ coordinates are ticked by arcseconds. \label{fig2}}
\end{figure}

\begin{figure}
\includegraphics[angle=0,scale=.60]{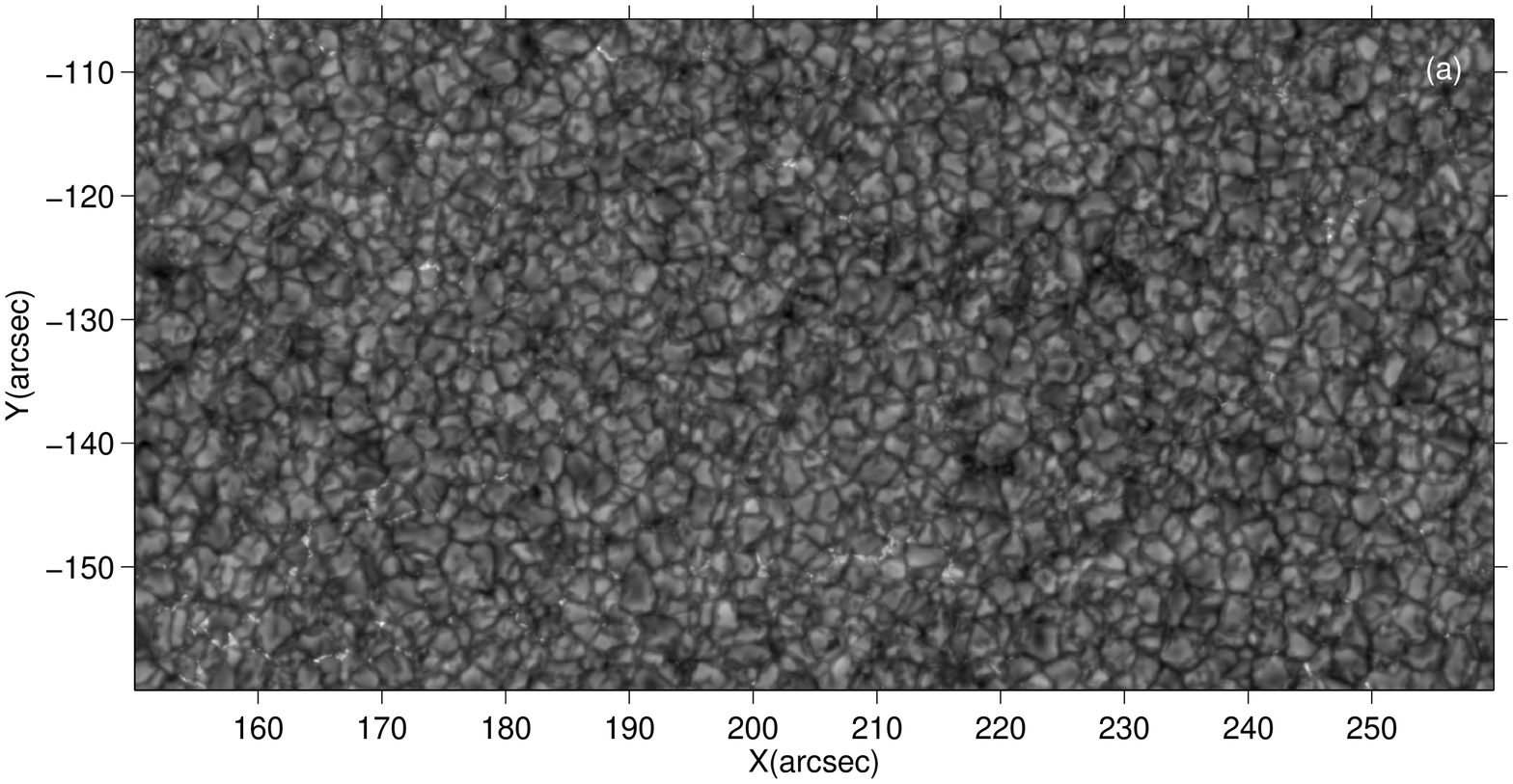}
\\
\\
\includegraphics[angle=0,scale=.666]{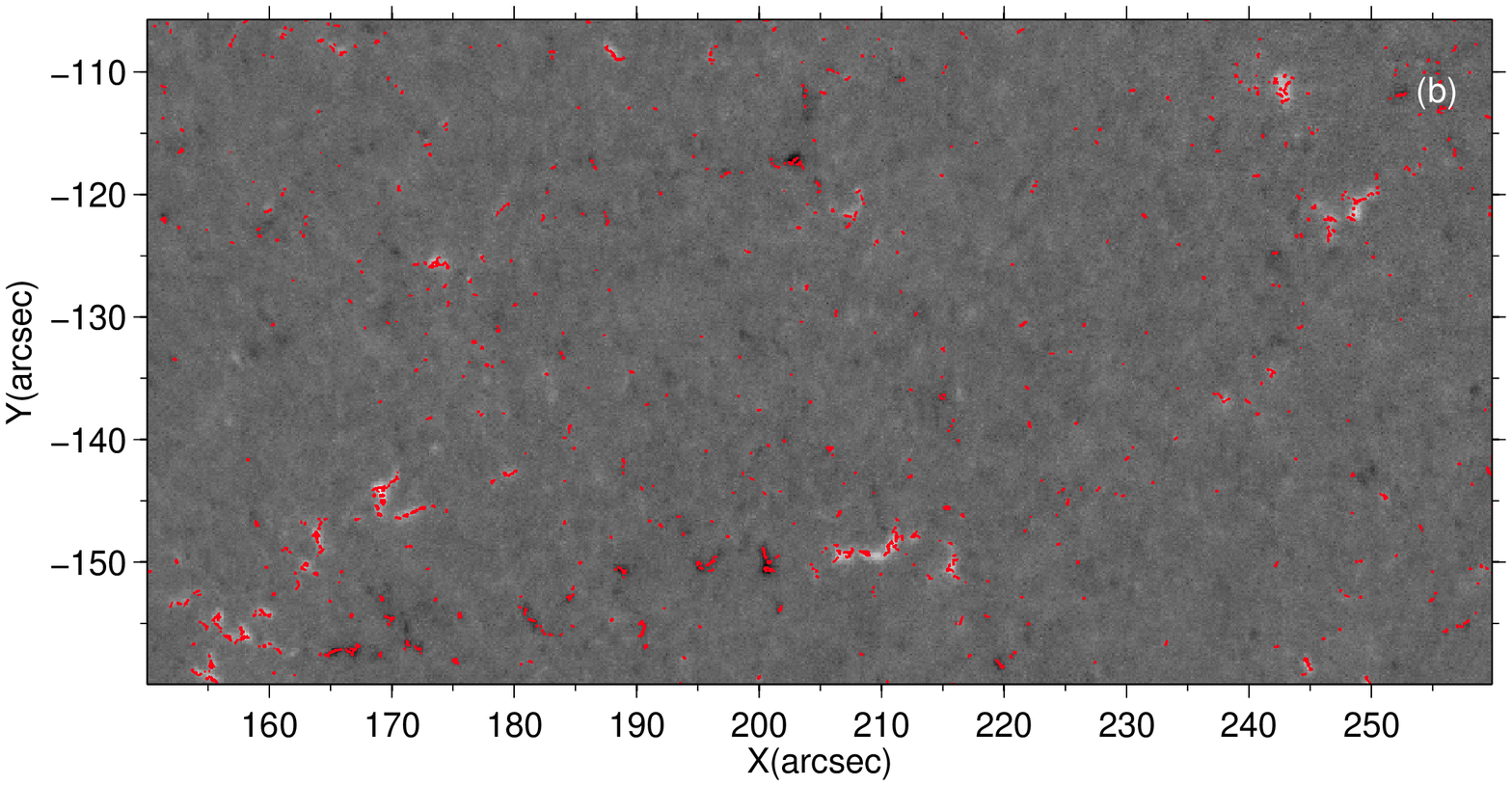}
\caption{Panel (a): a G-band image of data set \uppercase\expandafter{\romannumeral2} (a quiet Sun region), which the size is 109.8$''\times$54.3$''$. Panel (b): the co-aligned NFI magnetogram; the GBPs identified from the G-band image are highlighted with red. The $x$ and $y$ coordinates are ticked by arcseconds. \label{fig3}}
\end{figure}

A Laplacian and morphological dilation algorithm \citep{Feng13} was used to detect GBPs in each G-band image, and then a three-dimensional segment algorithm \citep{Yang14} was employed to track the evolution of GBPs in the timeseries of G-band images. Figure~\ref{fig2} shows a G-band image of data set \uppercase\expandafter{\romannumeral1} and the corresponding NFI magnetogram, in which the positions of the GBPs are highlighted in red. The GBPs cover 3.5\% of the selected FOV. Figure~\ref{fig3} shows the images of data set \uppercase\expandafter{\romannumeral2}, but the GBPs only cover 1\%. \citet{Sanchez04} proposed that the GBPs cover 0.7\% in the internetwork regions. Later, \citet{Sanchez10} detailed that the value was between 0.9\% and 2.2\% in QS regions. Recently, the fractional area occupied by GBPs in ARs is measured as twice to triple larger than that in QS regions \citep{Romano12, Feng13}.

For reducing detection error, these GBPs were discarded if (1) their equivalent diameters are less than 100\,$\rm km$ or greater than 500\,$\rm km$, (2) their lifetimes are shorter than 60\,$\rm s$, (3) their horizontal velocities exceed 7\,$\rm km s^{-1}$, (4) their lifecycles are not complete, or (5) they merge or split during their lifetimes. As a result, a total of 103,023 bright points remain in 132 images, yielding 18,494 evolving GBPs of data set \uppercase\expandafter{\romannumeral1}; 19,349 bright points remain in 110 images, yielding 3,242 evolving GBPs in data set \uppercase\expandafter{\romannumeral2}.

\subsection{Extraction of the Magnetic field Strengths of GBPs}
The peak longitudinal magnetic field strength of each bright point was extracted from the corresponding NFI magnetogram after its region was identified. The reason why a peak value was adopted rather than the mean or median is that the scales of GBPs are so small that slight errors in image alignment can degrade average or median values significantly \citep{Berger07}.

The absolute strongest magnetic field strength, $B$, during the evolution of each GBP was calculated because it represents the peak state. Taking the noise of the NFI magnetogram into account, we discarded the GBPs from data set \uppercase\expandafter{\romannumeral1} and \uppercase\expandafter{\romannumeral2} with a $B$ are below 50 and 100\,$\rm G$ (about 4$\sigma$), respectively. About 97\% GBPs of data set \uppercase\expandafter{\romannumeral1} fall into a range of 50\,--\,1000\,$\rm G$, while 97.5\% of data set \uppercase\expandafter{\romannumeral2} fall into 100\,--\,450\,$\rm G$. Consequently, these GBPs of data set \uppercase\expandafter{\romannumeral1} and \uppercase\expandafter{\romannumeral2} were categorized into 19 and 7 groups by setting the same bin as 50\,$\rm G$, respectively.

\subsection{Diffusion Approaches}
A traditional analysis of the diffusion process is the Lagrangian approach, which is efficient when we deal with diffusive properties of tracers within turbulent fluid flows \citep{Monin75}. The approach consists of two main steps: (1) computing the spatial displacement, $\triangle l$, of an individual GBP as a function of time interval, $\tau$, measured from its first appearance; (2) calculating the mean-square displacement, $\langle(\triangle l)^{2}\rangle$, for each time interval as a function of $\tau$. The power index, $\gamma$, of the mean-square displacement is defined as
\begin{equation}\label{eqa1}
\langle(\triangle l)^{2}\rangle = C\tau^{\gamma},
\end{equation}
where, $C$ is the coefficient of proportionality. Usually, $\gamma$ and $C$ are derived as the slope and the exponential of the intercept on the $y$ axis of the spectrum over a range of $\tau$ on a log-log scale, respectively. However, this approach is not ideal for studying the random motions of GBPs \citep{Dybiec09, Jafarzadeh14}. These authors indicated that the second step might cause the mixing of different diffusive processes. Thus, they suggested that the square displacement of each GBP ought to be calculated separately. The $\gamma$ value could be measured by the slope of its own square displacement on a log-log scale. Then, the mean diffusion index and the associated standard deviation (the square root of the variance) could be obtained by fitting the distribution of the $\gamma$ values. We named this approach distribution of diffusion indices (DDI).
In this study, both of the approaches were adopted and compared.

\section{RESULT}
\subsection{Lagrangian Approach}
\begin{figure}
\includegraphics[angle=0,scale=.6]{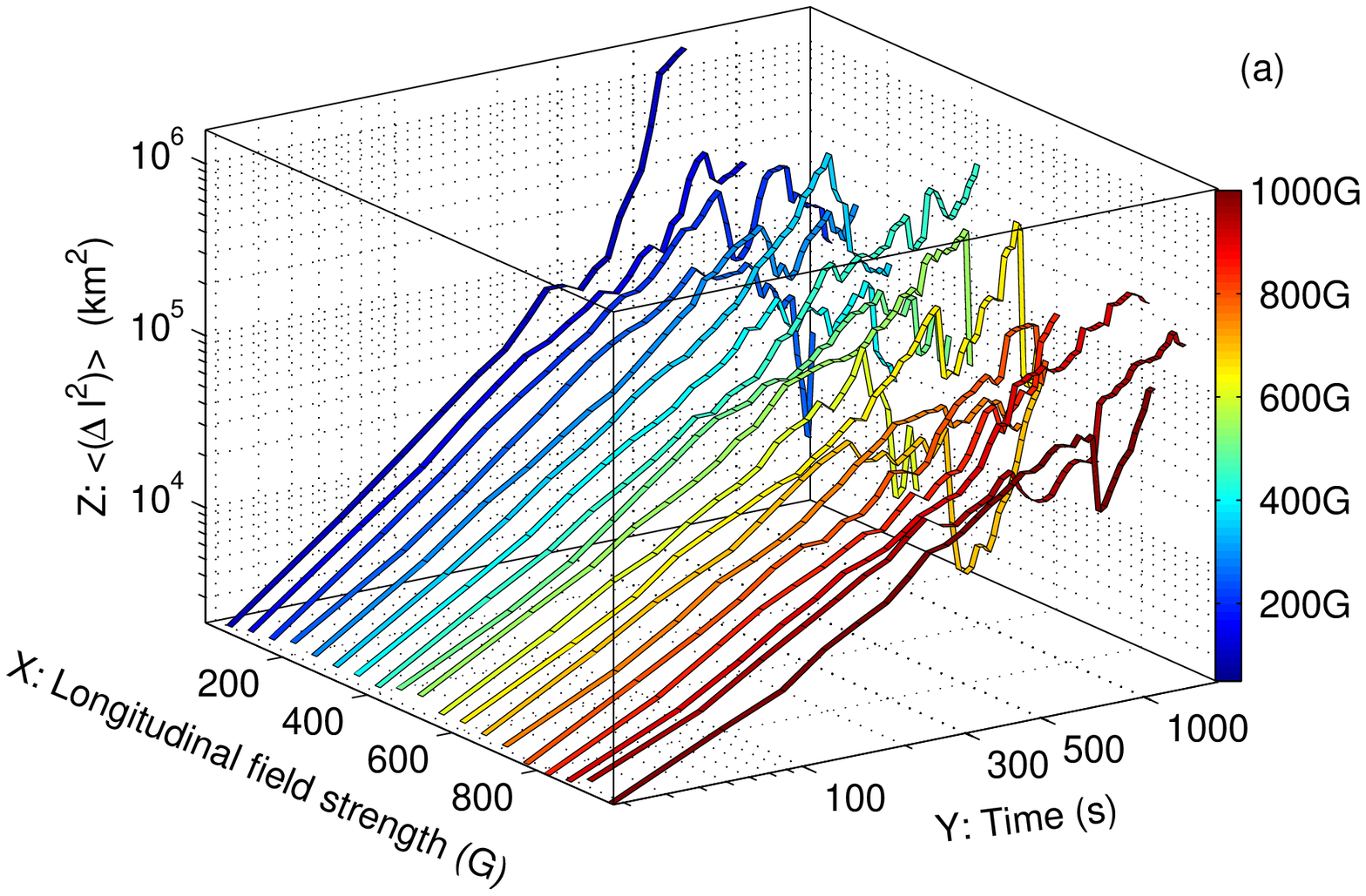}
\\
\\
\includegraphics[angle=0,scale=.6]{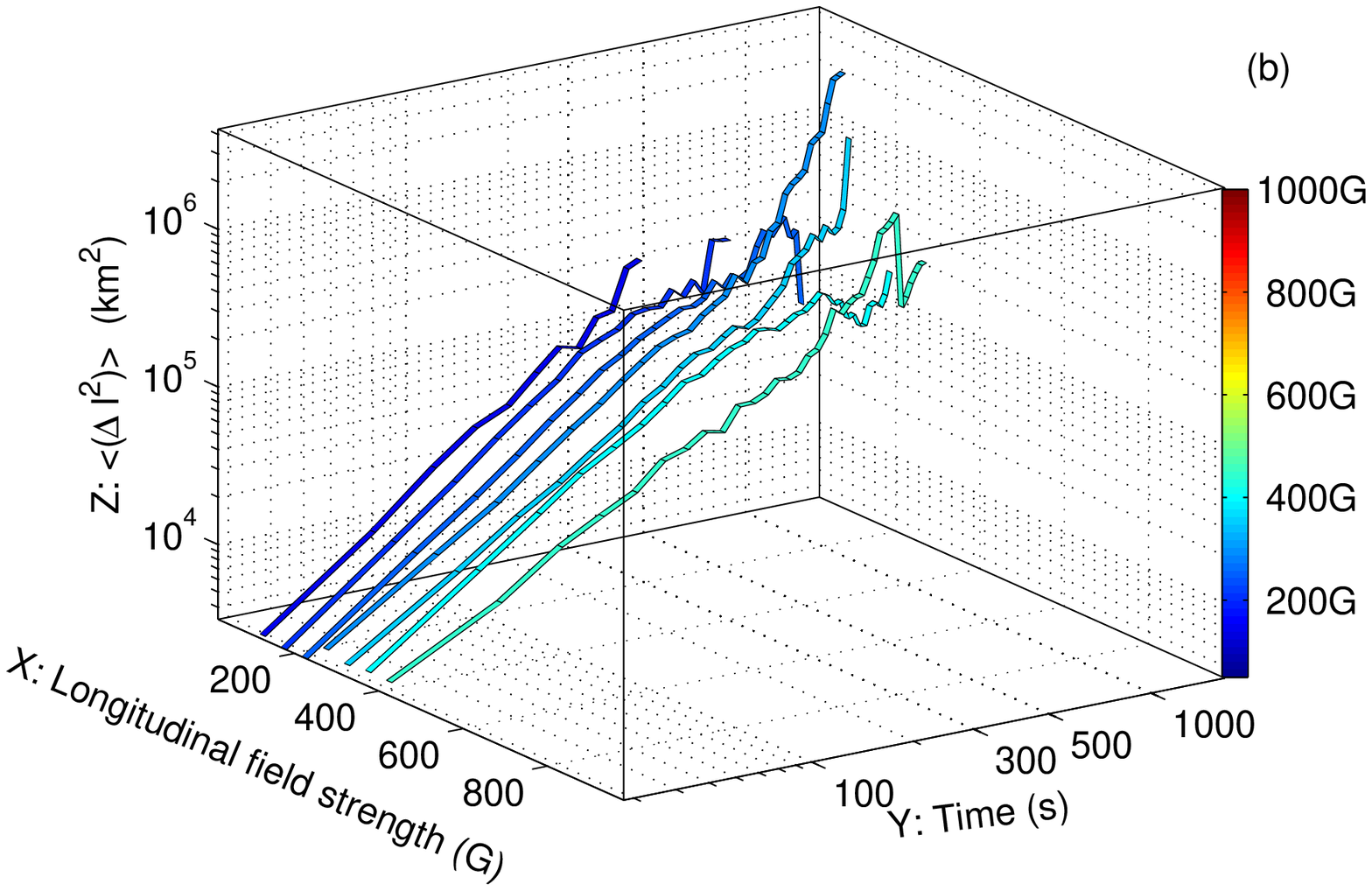}
\caption{The mean-square displacement $\langle(\triangle l)^{2}\rangle$ of GBPs as a function of time, $\tau$, on a log-log scale in different longitudinal magnetic field strength bins by the Lagrangian approach. Panel (a): The mean-square displacement of data set \uppercase\expandafter{\romannumeral1}, which the longitudinal magnetic field strength bins range from 50 to 1000\,$\rm G$. Panel (b): The mean-square displacement of data set \uppercase\expandafter{\romannumeral2}, which the longitudinal magnetic field strength bins range from 100 to 450\,$\rm G$.\label{fig4}}
\end{figure}

The mean-square displacements of GBPs in different longitudinal magnetic field strength bins versus the times of data set \uppercase\expandafter{\romannumeral1} and \uppercase\expandafter{\romannumeral2} are displayed on a log-log scale in Figure~\ref{fig4} (a) and (b), respectively.
Many previous authors have proposed that the estimation of $\gamma$ using the Lagrangian approach may be strongly affected by timescales. In particular, \citet{Abramenko11} interpreted that the $\gamma$ value measured for small timescales ($\lesssim$ 300\,$\rm s$) could represent the intrinsic diffusion of GBPs. As shown in Figure~\ref{fig4}, the lengths of mean-square displacements are different and the tails are not regular because only a few GBPs have long lifetimes. If the $\gamma$ value was calculated by fitting the whole mean-square displacement, it would be determined by a few long-lived GBPs. Therefore, we analyzed the mean-square displacements for the timescale $\tau$\,$\lesssim$ 300\,$\rm s$ to probe the intrinsic depiction of the diffusion index. The $\gamma$ values and the associated standard deviations were conveniently captured, respectively, by the slopes of mean-square displacements for the timescale $\tau$\,$\lesssim$ 300\,$\rm s$ on a log-log scale within 95\% confidence intervals.

\begin{figure}
\epsscale{1.0}
\plottwo{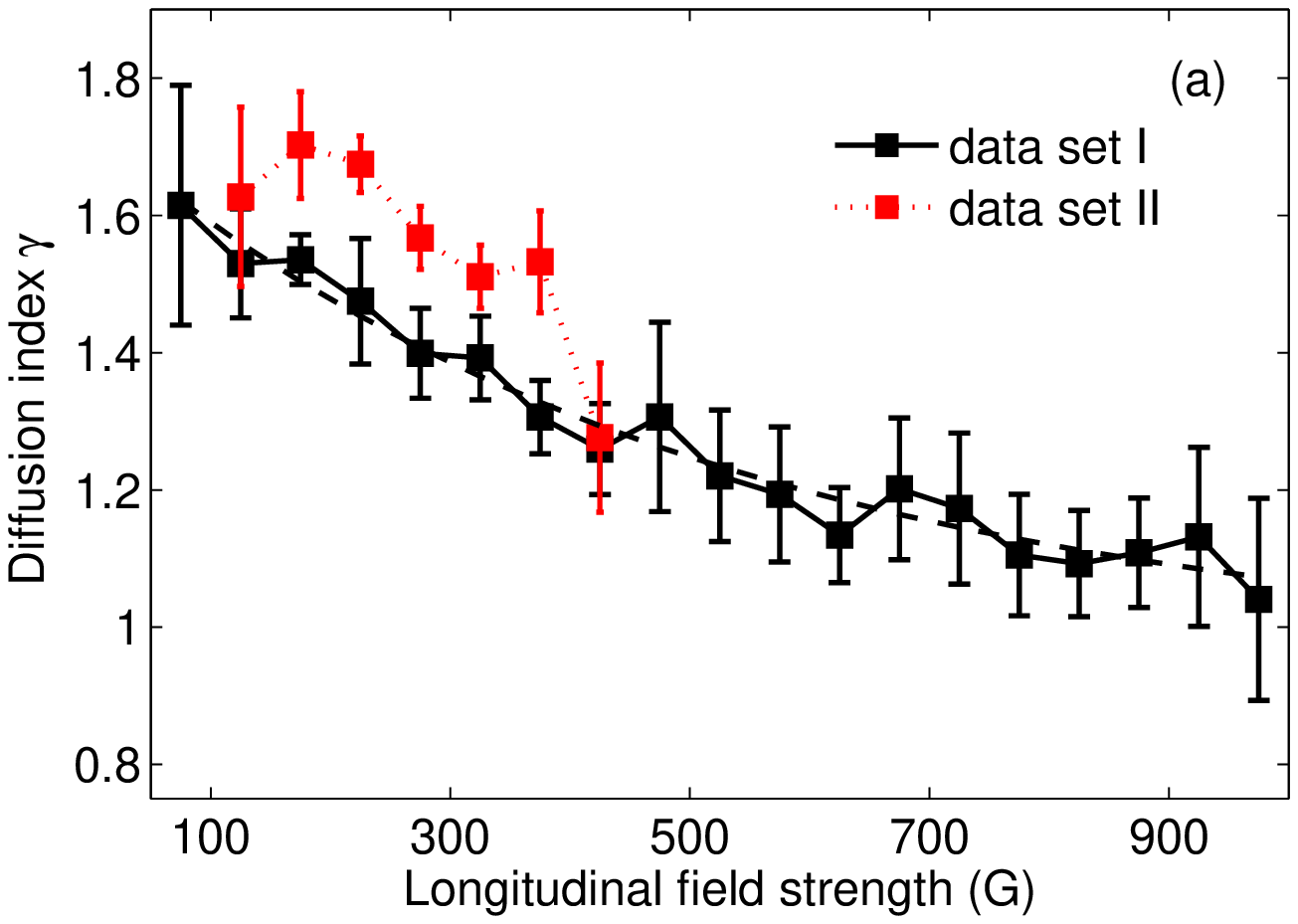}{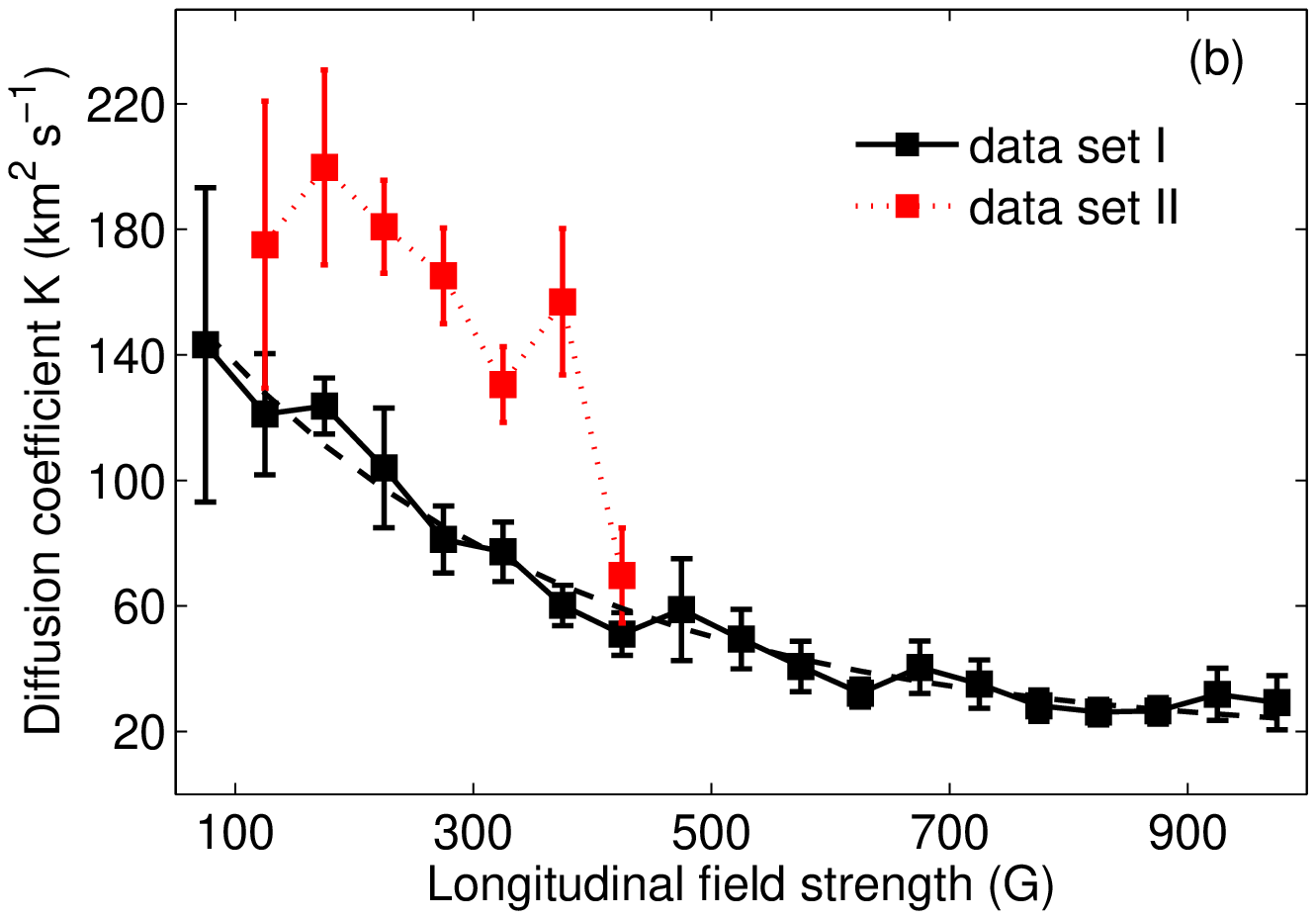}
\caption{Panel (a): the relations between the diffusion indices of GBPs and the longitudinal magnetic field strengths for the timescale is less than 300\,$\rm s$ by the Lagrangian approach. Panel (b): the relations between the diffusion coefficients of GBPs and the longitudinal magnetic field strengths for the timescale is less than 300\,$\rm s$ by the Lagrangian approach. The mean values and the standard deviations of data set \uppercase\expandafter{\romannumeral1} are illustrated using the error bars in black solid lines, which are fitted well with an exponential function (black dashed line). The relation of data set \uppercase\expandafter{\romannumeral2} is illustrated with a red dotted line.\label{fig5}}
\end{figure}

In Figure~\ref{fig5} (a), the mean diffusion indices and the associated standard deviations of the different longitudinal magnetic field strength bins of data set \uppercase\expandafter{\romannumeral1} are illustrated using error bar in black solid line. Taking the quantity of each bin as a weight, we analyzed the relation between the diffusion indices and the longitudinal magnetic field strengths using a weighted curve fitting. It fits well with an exponential function (black dashed line). The $\gamma$ value and the standard deviation are 1.61$\pm$0.17 (super-diffusion) inside the bin of 50\,--\,100\,$\rm G$. As the longitudinal magnetic field strength increases, the $\gamma$ value decreases. At a strong longitudinal magnetic field strength, the gradient becomes small and the $\gamma$ value arrives at $\sim$1.00 (normal-diffusion). The empirical formula deduced from the fitting equation is given to be:
\begin{equation}\label{eqa2}
\hat{\gamma}(B)= a  e^{b B} + c,
\end{equation}
where $B$ is the longitudinal magnetic field strength in kG. The parameters $a$, $b$, and $c$ are 0.77$\pm$0.10, -1.95$\pm$0.91, and 0.96$\pm$0.15 under the 95\% confidence interval, respectively.

The diffusion indices of data set \uppercase\expandafter{\romannumeral2} are also shown in Figure~\ref{fig5} (a) with a red dotted line. Compared with data set \uppercase\expandafter{\romannumeral1}, the trend is simple because the longitudinal magnetic field strengths of GBPs only range from 100 to 450\,$\rm G$. The $\gamma$ value continues to mostly decrease from 1.70$\pm$0.07 to 1.27$\pm$0.10, except the first $\gamma$ value is smaller than the second, and the fifth is slightly smaller than the sixth. In Figure~\ref{fig4} (b), it can be seen that a sudden drop of the mean-square displacement happens at $\tau \simeq$ 300\,$\rm s$ in the first longitudinal field strength bin. This leads to a small $\gamma$ value with a large standard deviation. Because of the limited range of longitudinal magnetic field strengths, we neglected the fitting formula.

We then established the diffusion coefficient, $K$, of anomalous diffusion with the formula \citep{Monin75}:
\begin{equation}\label{eqa3}
K(\tau)= \frac{C\gamma}{4}\tau^{\gamma-1},
\end{equation}
where $\gamma$, $\tau$, and $C$ are deduced from equation (\ref{eqa1}). Figure~\ref{fig5} (b) shows the relations between $K$ and $B$ for the timescale $\tau\,\lesssim$ 300\,$\rm s$. The $K$ value of data set \uppercase\expandafter{\romannumeral1} decreases exponentially from 143$\pm$50 to 26$\pm$4\,$\rm km^{2}$\,$\rm s^{-1}$. The empirical formula is given as:
\begin{equation}\label{eqa4}
\hat{K}(B)= a  e^{b B} + c,
\end{equation}
where $B$ is the longitudinal magnetic field strength in kG. The parameters $a$, $b$, and $c$ are 165.30$\pm$15, -3.23$\pm$0.87, and 17.22$\pm$12 under 95\% confidence interval, respectively.
The trend of $K$ of data set \uppercase\expandafter{\romannumeral2} is similar to the corresponding $\gamma$. The $K$ value decreases from 200$\pm$31 to 69$\pm$15\,$\rm km^{2}$\,$\rm s^{-1}$ except the first $K$ value is smaller than the second, and the fifth is also smaller than sixth. It can be seen that the $K$ values are distinctly higher than those of data set \uppercase\expandafter{\romannumeral1}. In equation (\ref{eqa1}) and (\ref{eqa3}), $\gamma$ relates to the acceleration and $C$ relates to the initial velocity. The $C$ values deduced from equation (\ref{eqa1}) of data set \uppercase\expandafter{\romannumeral2} are greater than those of \uppercase\expandafter{\romannumeral1}. It is in agreement with the previous studies, where the horizontal velocity of GBPs is attenuated in ARs compared to QS regions \citep{Berger98, Mostl06, Keys11}. This is the main reason why the $K$ values of data set \uppercase\expandafter{\romannumeral2} are distinctly higher than those of data set \uppercase\expandafter{\romannumeral1}.

\subsection{DDI Approach}
\begin{figure}
\includegraphics[angle=0,scale=.6]{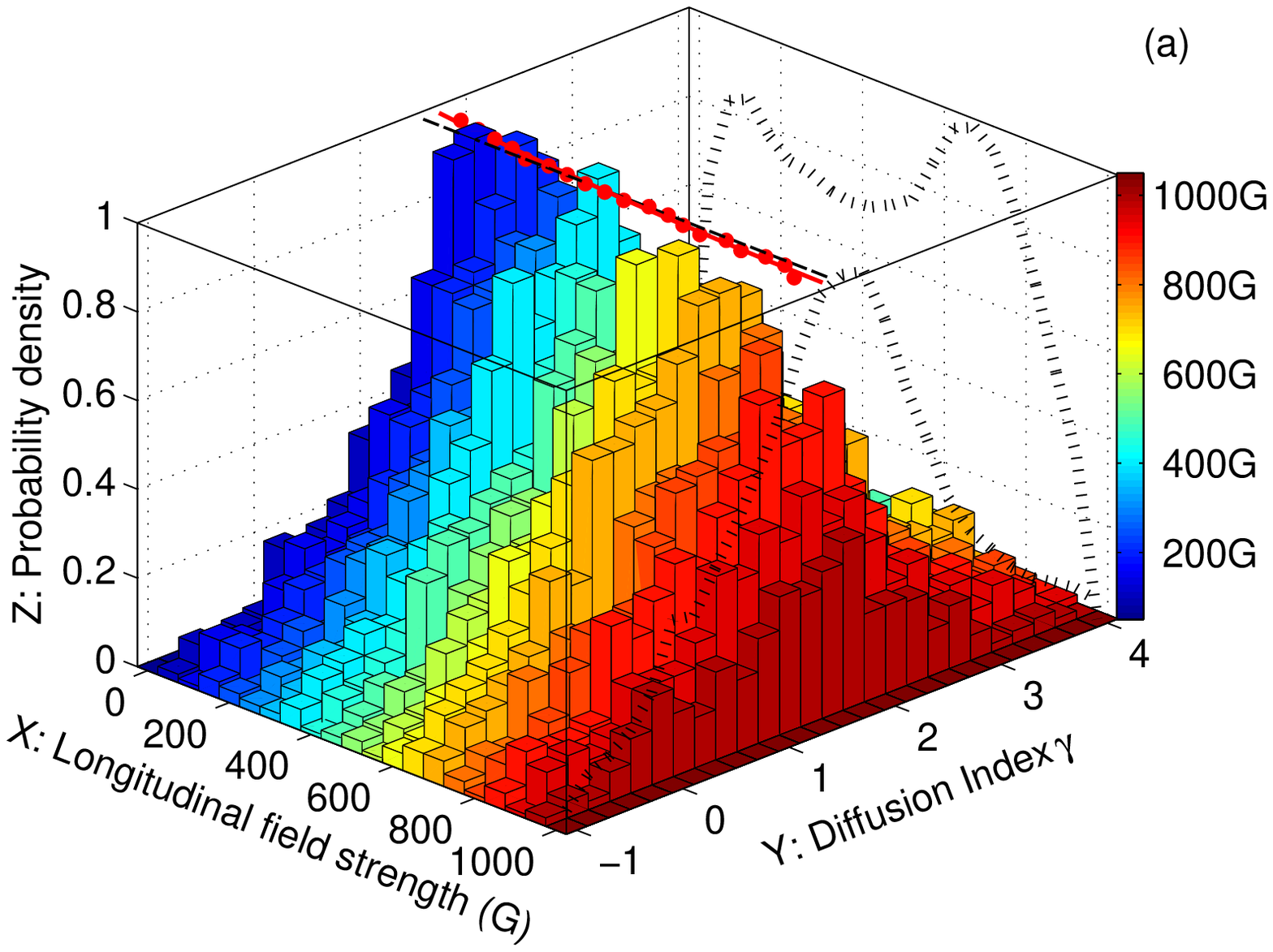}
\\
\\
\includegraphics[angle=0,scale=.6]{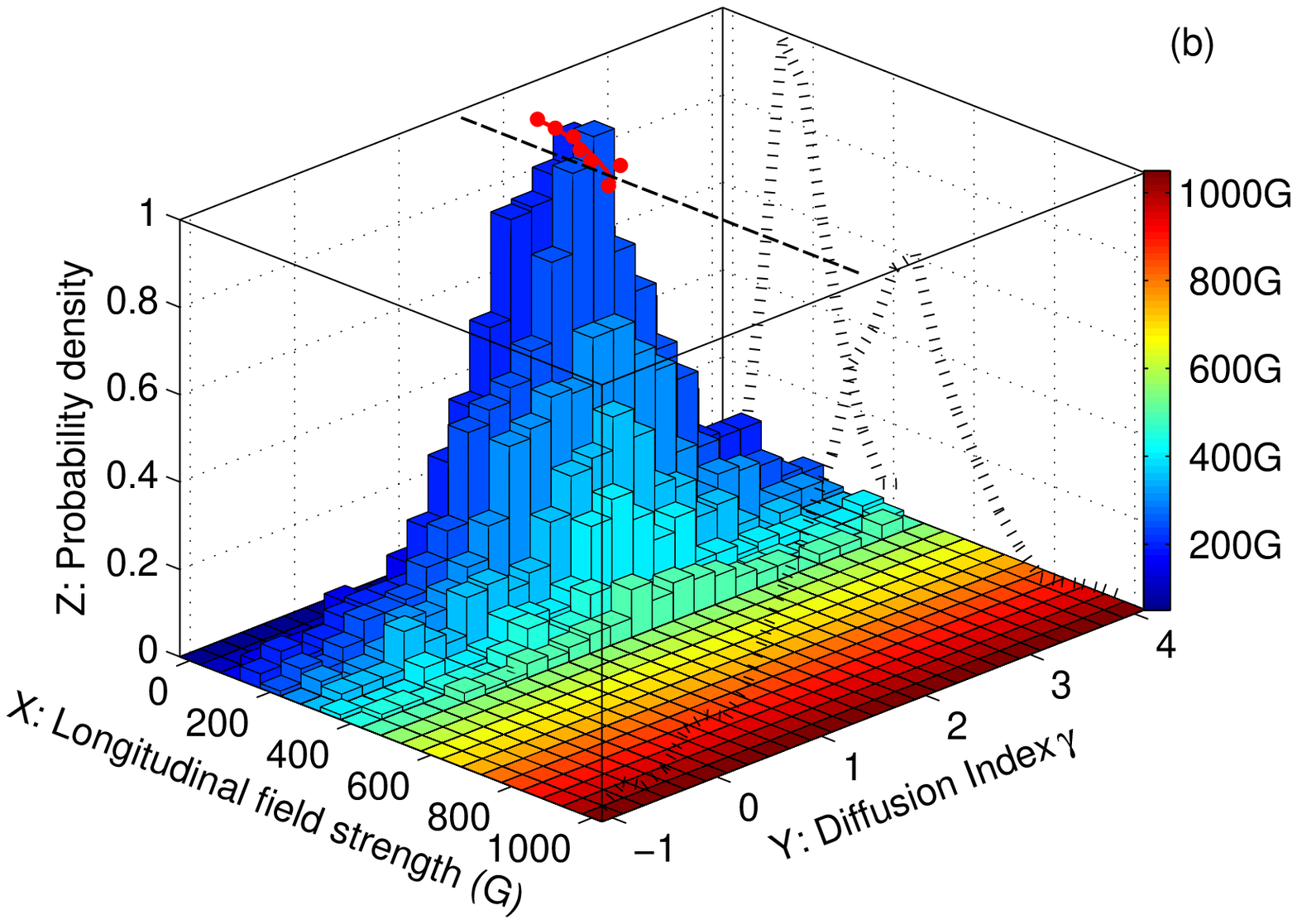}
\caption{The two-dimensional histograms of the diffusion indices of GBPs and the longitudinal magnetic field strengths by the DDI approach. The marginal distribution of the longitudinal magnetic field strengths is projected on the $xoz$ plane. The marginal distribution of the diffusion indices is projected on the $yoz$ plane. On the upper $xoy$ plane, the mean $\gamma$ values of all bins, and a quadratic curve fit of the mean $\gamma$ values are drawn in red. Panel (a): the histogram of data set \uppercase\expandafter{\romannumeral1}. Panel (b): the histogram of data set \uppercase\expandafter{\romannumeral2}. \label{fig6}}
\end{figure}

Figure~\ref{fig6} (a) shows the relation between $\gamma$ and $B$ in data set \uppercase\expandafter{\romannumeral1} using the DDI approach in the two-dimensional histogram. The $\gamma$ value of each GBP is measured by the slope of its own square displacement and lifetime on a log-log scale. The distribution of $\gamma$ in each bin is fitted with a Gaussian function well and the marginal distribution of $\gamma$ is projected on $yoz$ plane. All distributions have similar shapes, but shift with the increasing longitudinal field strengths. The ranges of $\gamma$ values in different bins have no significant difference. About 83\%$\sim$90\% $\gamma$ values range from 0 to 4. The mean $\gamma$ and the associated standard deviation are calculated by fitting the Gaussian distribution of each bin. We find that the mean $\gamma$ values continue to mostly decrease from 1.66 to 1.33, and the associated standard deviations limit in a small range from 0.04 to 0.07. At the first longitudinal field strength bin, the $\gamma$ value is 1.66$\pm$0.05, then it decreases with the increasing longitudinal field strength, and finally reaches a value of 1.33$\pm$0.07. On the upper $xoy$ plane, the mean $\gamma$ values of all bins and a quadratic curve fit of the mean $\gamma$ values are drawn in red.

Figure~\ref{fig6} (b) shows the relation between $\gamma$ and $B$ in data set \uppercase\expandafter{\romannumeral2}. The longitudinal field strengths range from 100 to 450\,$\rm G$. The distribution of $\gamma$ in each bin is also fitted well with a Gaussian function. In different bins, about 90\%$\sim$95\% $\gamma$ values range from 0 to 4. The $\gamma$ values decrease from 1.84$\pm$0.07 to 1.37$\pm$0.14, except for the value of 1.68$\pm$0.07 in the sixth bin.

\begin{figure}
\epsscale{1.0}
\plottwo{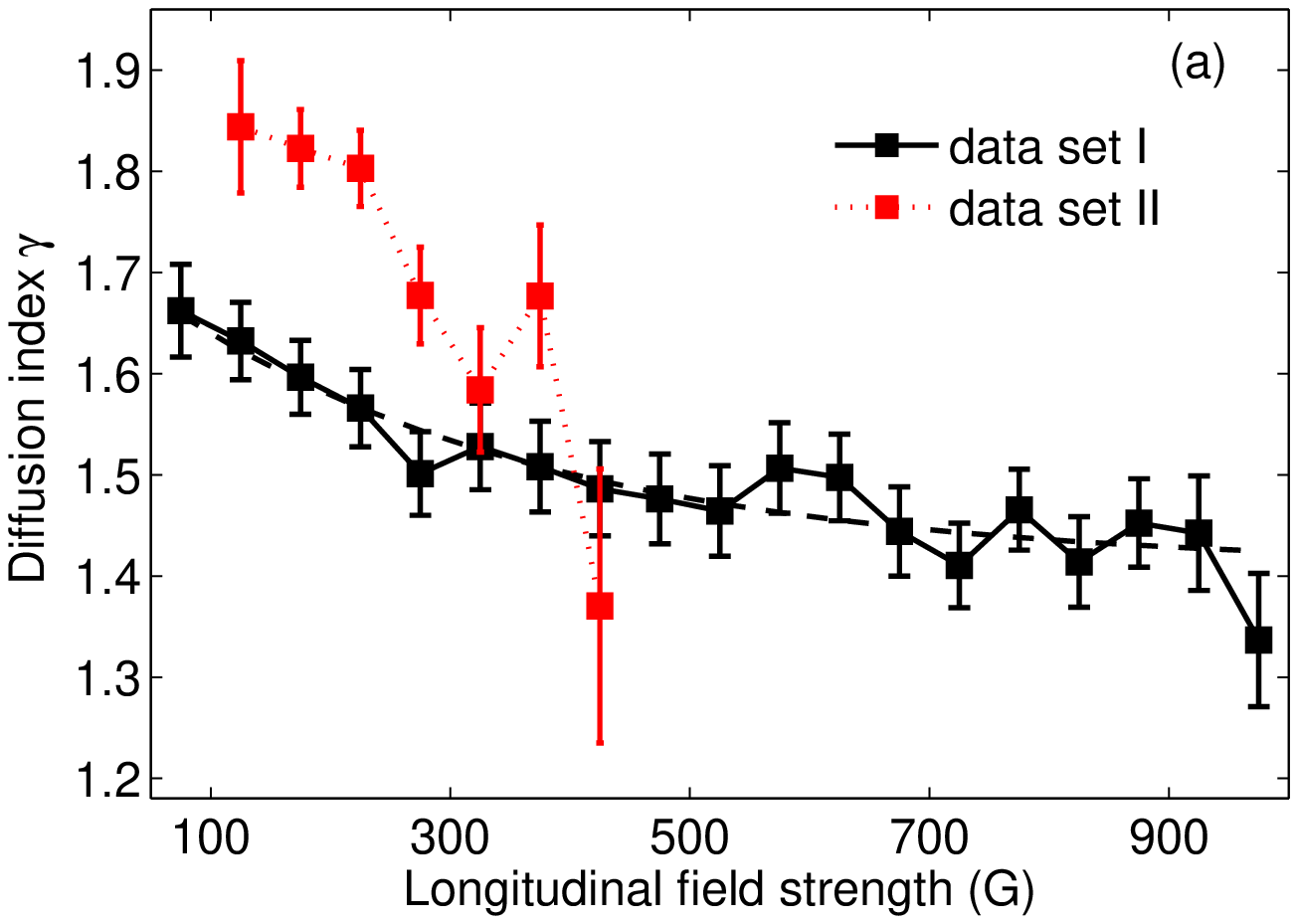}{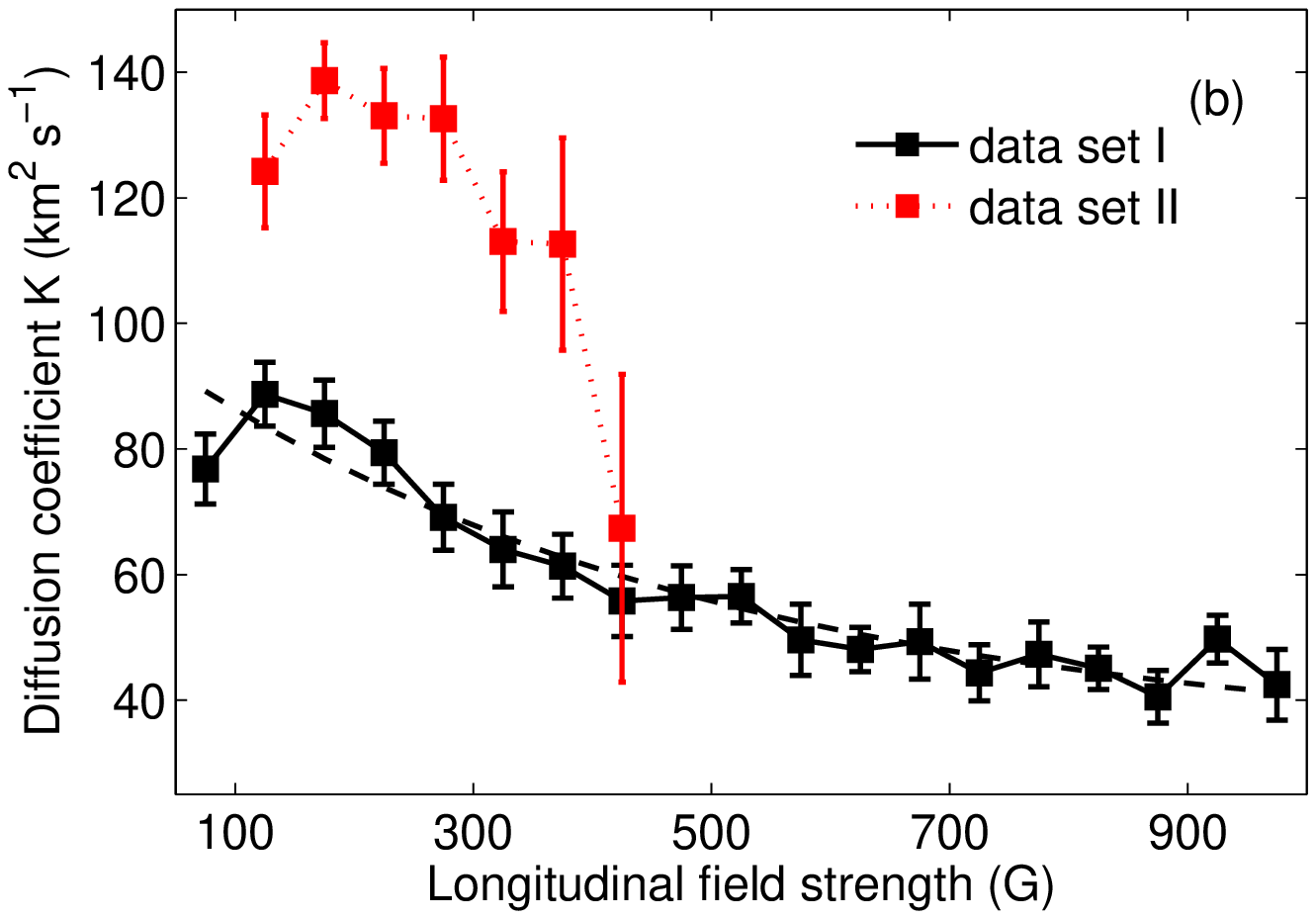}
\caption{Panel (a): the relations between the diffusion indices of GBPs and the longitudinal magnetic field strengths by the DDI approach using error bars. Panel (b): the relation between the diffusion coefficients of GBPs and the longitudinal magnetic field strengths. The mean values and the associated standard deviations of data set \uppercase\expandafter{\romannumeral1} and \uppercase\expandafter{\romannumeral2} are drawn with a black solid line and red dotted line, respectively. Both relations of data set \uppercase\expandafter{\romannumeral1} are fitted very well with exponential functions (black dashed lines). \label{fig7}}
\end{figure}

To explore the relation between $\gamma$ and $B$ more clearly, we redrew the mean $\gamma$ values and the associated standard deviations of data set \uppercase\expandafter{\romannumeral1} and \uppercase\expandafter{\romannumeral2} using the error bars in Figure~\ref{fig7} (a) with a black solid line and red dotted line, respectively. The trend of the mean $\gamma$ values of data set \uppercase\expandafter{\romannumeral1} is fitted well with an exponential function (black dashed line). By weighted curve fitting, the empirical formula was deduced as equation (\ref{eqa2}), where the parameters $a$, $b$ and $c$ are 0.32$\pm$0.07, -2.15$\pm$1.58, and 1.41$\pm$0.06 under the 95\% confidence interval, respectively.

Note that the marginal distributions of $B$ of data set \uppercase\expandafter{\romannumeral1} and \uppercase\expandafter{\romannumeral2} are projected on the $xoz$ plane in Figure~\ref{fig6} (a) and (b), respectively, which are fitted with a double log-normal distribution with two peaks at 214$\pm$72 and 662$\pm$91\,$\rm G$, and a log-normal distribution with a peak at 277$\pm$49\,$\rm G$. Interestingly, there are two peaks in data set \uppercase\expandafter{\romannumeral1}, but only one peak in data set \uppercase\expandafter{\romannumeral2}. The first peak value of data set \uppercase\expandafter{\romannumeral1} is close to the peak value of data set \uppercase\expandafter{\romannumeral2}.

The $K$ value of each GBP is calculated by equation (\ref{eqa3}) using its lifetime as the timescale $\tau$. The mean $K$ values are determined by the distributions of the $K$ values of different bins, respectively. Figure~\ref{fig7} (b) shows both relations between $K$ and $B$ of data set \uppercase\expandafter{\romannumeral1} and \uppercase\expandafter{\romannumeral2}. The $K$ value of data set \uppercase\expandafter{\romannumeral1} decreases exponentially from 89$\pm$5 to 41$\pm$4\,$\rm km^{2}$\,$\rm s^{-1}$. The empirical formula follows equation (\ref{eqa4}), where $a$ is 66.21$\pm$11.01, $b$ is -1.57$\pm$0.94, and $c$ is 32.80$\pm$15.32 under 95\% confidence interval. The $K$ value of data set \uppercase\expandafter{\romannumeral2} decreases from 139$\pm$6 to 67$\pm$24\,$\rm km^{2}$\,$\rm s^{-1}$ except for the value of 124$\pm$9\,$\rm km^{2}$\,$\rm s^{-1}$ inside the first bin. We find that the $K$ values in the first bins of both data sets are small, although the corresponding $\gamma$ values are not. The main reason is that the lifetimes of GBPs with weak field strengths are shorter than those with strong ones.

\section{DISCUSSION}
We used high spatial and temporal resolution G-band images and simultaneous NFI Stokes $I$ and $V$ images acquired with Hinode /SOT. The corresponding longitudinal magnetic field strength of each GBP was extracted from the calibrated NFI magnetogram after carefully aligning these data with the G-band images. The point-to-point method is feasible to investigate the diffusion regimes of magnetic flux tubes at different longitudinal magnetic strengths.

\subsection{Lagrangian and DDI Approach}
The Lagrangian approach and the DDI approach have been adopted to measure the $\gamma$ and $K$ values separately. The relations between $\gamma$ and $B$ and between $K$ and $B$ are both fitted well with exponential functions no matter which approach was used, although the values are somewhat different.

The traditional Lagrangian approach is typically used to analyze the diffusion of tracers within fluid flows. The step of mean-square displacement averages the displacement of individual GBPs, and then results in diminished displacement. It is the main reason that the $\gamma$ values are smaller than those of the DDI approach. In addition, this approach inevitably depends on timescales. A short timescale generally results in a small diffusion index, and vice versa. Most GBPs have a short lifetime because the distribution of lifetime follows an exponential function and the mean lifetime is about 150\,$\rm s$. Therefore, a few random long-lived GBPs will determine the diffusion efficiency for long timescales, which is illustrated in Figure~\ref{fig4}. The timescale was determined variously in previous studies. Some authors took the whole mean-square displacement directly, while others cut part of the tail of mean-square displacement based on the percentage or the goodness-of-fit, etc. We calculated the mean-square displacements of all longitudinal field strength bins for the timescale $\lesssim$ 300\,$\rm s$ because this timescale is not affected by a few long-lived GBPs.

Another approach is the DDI, where the mean diffusion index is obtained from the distribution of diffusion indices of individual GBPs. We analyzed all trajectories of individual GBPs separately. The approach avoids mixing GBPs in different diffusion regimes, so it sheds light on the intrinsic property of their proper motions. However, if a GBP moves in an erratic or circular path, the slope of linear fit on a log-log scale would be very small (even below zero) or very large (greater than four) with a low goodness-of-fit. Examples of such pathological cases have been given by \citet{Jafarzadeh14}. These cases result in meaningless diffusion indices. The percentages of meaningless diffusion indices of data set \uppercase\expandafter{\romannumeral1} and \uppercase\expandafter{\romannumeral2} are 13\% and 8\%, respectively. In detail, the very small diffusion indices (below zero) and very large (greater than four) account for 9.6\% and 3.5\% of data set \uppercase\expandafter{\romannumeral1}, 6\% and 2\% of data set \uppercase\expandafter{\romannumeral2}, respectively. Besides that, the $K$ value of each GBP is calculated using its lifetime as the timescale. That means most $K$ values are calculated by very short lifetimes because the lifetime distribution of GBPs follows an exponential function. According to equation (\ref{eqa3}), a shorter timescale will give rise to a smaller $K$ value. This directly leads to the $K$ values are smaller than those calculated for the timescale $\tau\,\lesssim$ 300\,$\rm s$ by the Lagrangian approach.

Above all, each of the two approaches has advantages and disadvantages respectively. We prefer to adopt the DDI approach to estimate the diffusion efficiency of GBPs because it reflects the diffusion regimes of individual GBPs.

\subsection{Diffusion Index and Diffusion Coefficient}

The $\gamma$ and $K$ values decrease with the increasing longitudinal magnetic field strengths, and they decrease exponentially of data set \uppercase\expandafter{\romannumeral1}. It is suggested that in the same environment, strong magnetic elements diffuse less than weak elements. In addition, Figure~\ref{fig5} and Figure~\ref{fig7} indicate that the $\gamma$ and $K$ values of GBPs in strong magnetized environments are less than those with the same longitudinal field strengths in weak ones. The $\gamma$ values of data set \uppercase\expandafter{\romannumeral1} and \uppercase\expandafter{\romannumeral2} by the Lagrangian approach are 1.30$\pm$0.09 and 1.61$\pm$0.06 for the timescale  $\tau$\,$\lesssim$ 300\,$\rm s$, and by the DDI approach they are 1.53$\pm$0.01 and 1.79$\pm$0.01, respectively. The $K$ values using the Lagrangian approach are 56$\pm$14 and 165$\pm$41\,$\rm km^{2}$\,$\rm s^{-1}$  for $\tau$\,$\lesssim$ 300\,$\rm s$, and are 78$\pm$29 and 130$\pm$54\,$\rm km^{2}$\,$\rm s^{-1}$, respectively, using the DDI approach.

Some authors analyzed the diffusion in special regions. In network regions of the QS or in ARs, \citet{Berger98} got $\gamma$\,=\,1.34$\pm$0.06. \citet{Cadavid99} found $\gamma$\,=\,0.76$\pm$0.04 for timescales shorter than 22 minutes and 1.10$\pm$0.24 for timescales longer than 25 minutes. \citet{Lawrence01} indicated $\gamma$\,=\,1.13$\pm$0.01. \citet{Wang98} reported $K$\,=\,150\,$\rm km^{2}$\,$\rm s^{-1}$. \citet{Schrijver96} argued that a corrected diffusivity of 600\,$\rm km^{2}$\,$\rm s^{-1}$ is in a good agreement with the well performing model for the magnetic flux transport during a solar cycle \citep{Wang94}. By tracking magnetic features in MDI magnetograms, \citet{Hagenaar99} found $K$\,=\,70\,--\,90\,$\rm km^{2}$\,$\rm s^{-1}$. The smallest $K$ value of 0.87\,$\rm km^{2}$\,$\rm s^{-1}$ was reported by \citet{Chae08} from Hinode /SOT NFI magnetograms. Based on solving the equation of magnetic induction, they modeled the difference of the magnetic field in individual pixels between two frames to measure the magnetic diffusivity. The speculated reason for such a small $K$ value is that their method aimed at individual pixels between two frames with a 10\,$\rm minute$ interval. In QS regions, \citet{Cameron11} described $K$ lying in the range of 100\,--\,340\,$\rm km^{2}$\,$\rm s^{-1}$. \citet{Chitta12} reported $\gamma$\,=\,1.59 and $K$=90\,$\rm km^{2}$\,$\rm s^{-1}$. \citet{Yang15} proposed $\gamma$ \,=\,1.50 and $K$=191$\pm$20\,$\rm km^{2}$\,$\rm s^{-1}$. \citet{Jafarzadeh14} obtained $\gamma$ of 1.69$\pm$0.08 and $K$=257$\pm$32\,$\rm km^{2}$\,$\rm s^{-1}$ with the Ca \textsc{ii} H data using the DDI approach.

Some authors also compared the diffusion in different regions. \citet{Schrijver90} found the $K$ values of 110 and 250\,$\rm km^{2}$\,$\rm s^{-1}$ in the core of a plage region and surrounding quiet regions, respectively. \citet{Berger98} determined 60$\pm$11\,$\rm km^{2}$\,$\rm s^{-1}$ for network GBPs by assuming normal diffusion ($K =\langle(\bigtriangleup l)\rangle^{2}/4\tau$). However, when they reconstructed the velocity field using correlation tracking techniques, they presented the $K$ values of 77, 176, and 285\,$\rm km^{2}$\,$\rm s^{-1}$ in the magnetic, network, and quiet regions, respectively. \citet{Abramenko11} used the TiO data obtained with the New Solar Telescope (NST) of the Big Bear Solar Observatory (BBSO). They indicated that $\gamma$ and $K$ are 1.48 and 12\,$\rm km^{2}$\,$\rm s^{-1}$ for an AR, 1.53 and 19\,$\rm km^{2}$\,$\rm s^{-1}$ for a QS region, and 1.67 and 22\,$\rm km^{2}$\,$\rm s^{-1}$ for a CH. \citet{Giannattasio14} defined a network area and an internetwork area in the QS using Hinode /SOT NFI magnetograms. They found $\gamma$\,=\,1.27$\pm$0.05 and 1.08$\pm$0.11 in the network area (at smaller and larger scales, respectively), and 1.44$\pm$0.08 in the internetwork area. They also found $K$ value ranges from 80 to 150\,$\rm km^{2}$\,$\rm s^{-1}$ in a network area, and from 100 to 400\,$\rm km^{2}$\,$\rm s^{-1}$ in an internetwork area. However, some authors had different opinions. \citet{Keys14} estimated $\gamma$ values of $\sim$1.2 and $K$ values of $\sim$120$\pm$45\,$\rm km^{2}$\,$\rm s^{-1}$ for three subfields of varying magnetic flux densities. They proposed that the diffusion regimes in all three subfields were the same. Their $\gamma$ and $K$ values are consistent with those other authors analyzed in network regions, and more like the saturation value that we found at a strong longitudinal magnetic field strength.

Previous works suggested that magnetized elements diffuse more slowly in strongly magnetized regions than in less magnetized ones, although the range of reported $\gamma$ and $K$ values are large. Such a large range may be due to different instruments, data, and methods, especially magnetic flux densities of environments. Our results agreed with their conclusions. The absence of strong magnetic fields in the medium makes it possible for the bright points to diffuse faster (perhaps because of less interactions), resulting in larger diffusion indices and diffusion coefficients. Importantly, we find that strong magnetic elements diffuse less than weak elements, both in strong magnetized environments and in weak ones.

\subsection{Diffusion Regime}

The magnetic field is ubiquitous in the photosphere and interacts with convective flows at different scales. Strongly magnetized elements are not easily perturbed by convective (supergranular, mesogranular, granular) motions. They withstand the perturbing action of convective flows much better than the weak magnetic elements, so their motions are slower and result in smaller diffusion indices. \citet{Giannattasio14} interpreted that magnetic elements are transported effectively by convection, where a weak-field regime holds, especially internetwork; whereas, in network, where is a strong field, magnetic elements cannot be further transported and concentrated due to reduced convection. \citet{Abramenko11} indicated that GBPs in strong magnetic fields are so crowded within narrow intergranular lanes when compared with a weak magnetic environment, where GBPs can move freely due to a lower population density.

Additionally, the difference has been also interpreted as the combined effect of granular, mesogranular, and supergranular flows. \citet{Spruit90} showed that mesogranular flows decrease when approaching mesogranular boundaries. \citet{Orozco12} found that magnetic elements in the internetwork start accelerating radially outward at the center of a supergranular cell, and decelerate as they approach the boundaries of supergranules. \citet{Jafarzadeh14} analyzed that this is caused by an increasing velocity with radial distance of the supergranular flow profile due to mass conservation assuming a constant upflow over most of supergranules. Supergranular flows systematically advect all GBPs toward the boundaries of supergranules, and granular and mesogranular flows impart the GBPs with additional velocity. The GBPs at the internal of supergranular cell tend to follow a super-diffusive regime because the motions of GBPs are imparted mainly by supergranular flow and intergranular turbulence. Once these GBPs reach the boundaries of supergranules, they decelerate because they are trapped in the sinks due to inflows from the opposite directions (i.e., from neighbouring supergranules). Granular motions are still active, however, causing GBPs to undergo normal or even sub-diffusion processes.

Note that, this study only involved isolated GBPs. Non-isolated GBPs, which are most likely located at stagnation points, were excluded due to the difficulty of measuring of their displacements. Therefore, sub-diffusive GBPs might be underestimated.

\section{CONCLUSION}
We have presented a study of the dispersal of GBPs at different longitudinal magnetic field strengths. Two different environments are considered, namely, a strongly magnetized AR and a weakly magnetized quiet Sun region, characterized by different mean longitudinal magnetic field strengths of 132 and 64\,$\rm G$, respectively. The corresponding data sets were acquired with Hinode /SOT comprising FG G-band filtergrams (BFI) and Stokes $I$ and $V$ images (NFI). After identifying and tracking GBPs in the G-band images, we extracted the corresponding longitudinal magnetic field strength of each GBP from the co-aligned and calibrated NFI magnetograms, then categorized the GBPs into different groups by their strongest longitudinal magnetic field strengths during their lifetimes. The Lagrangian approach and the distribution of indices (DDI) approach were adopted separately to explore the diffusion efficiency of GBPs in different longitudinal magnetic field strength groups. The values of the diffusion index by the Lagrangian approach are generally smaller than those by the DDI approach. The main reason is that the step of mean-square displacement in the Lagrangian approach averages the displacement of individual GBPs and results in diminished displacement.

We find that the values of the diffusion index and the diffusion coefficient both decrease exponentially with the increasing longitudinal magnetic field strengths. The empirical formulas deduced from exponential fitting and the parameters have been presented in Section 4. Stronger elements show lower diffusion indices and diffusion coefficients both in active regions and in QS regions. Additionally, the diffusion indices and coefficients are generally larger in regions of low mean magnetic flux (i.e., in the quiet Sun).

The different diffusion regimes of GBPs are mainly set by convections. Strongly magnetized elements are not easily perturbed by convective motions, and vice versa. The reason they have slower motions and smaller diffusion indices and diffusion coefficients is that they withstand the perturbing action of convective flows much better than weak ones. Their magnetic energy is not negligible compared with the kinetic energy of the gas, and therefore the flows cannot perturb them so easily.

\acknowledgments
The authors are grateful to the anonymous referee for constructive comments and detailed suggestions to this manuscript. The authors are grateful to the support received from the National Natural Science Foundation of China (No: 11303011, 11263004, 11463003, 11163004, U1231205), Open Research Program of the Key Laboratory of Solar Activity of the Chinese Academy of Sciences (No: KLSA201414, KLSA201309). This work is also supported by the Opening Project of Key Laboratory of Astronomical Optics \& Technology, Nanjing Institute of Astronomical Optics \& Technology, Chinese Academy of Sciences (No. CAS-KLAOT-KF201306) and the open fund of the Key Laboratory of Modern Astronomy and Astrophysics, Nanjing University, Ministry of Education, China. The authors are grateful to the $Hinode$ team for the possibility to use their data. Hinode is a Japanese mission developed and launched by ISAS/JAXA, collaborating with NAOJ as a domestic partner, NASA and STFC (UK) as international partners. Scientific operation of the $Hinode$ mission is conducted by the $Hinode$ science team organized at ISAS/JAXA. This team mainly consists of scientists from institutes in the partner countries. Support for the post-launch operation is provided by JAXA and NAOJ (Japan), STFC (U.K.), NASA (U.S.A.), ESA, and NSC (Norway).

\clearpage



\clearpage

\end{document}